\documentclass[fleqn,usenatbib]{mnras}

\usepackage{newtxtext,newtxmath}

\usepackage[T1]{fontenc}
\usepackage{ae,aecompl}

\usepackage{graphicx}	
\usepackage{amsmath}	
\usepackage{amssymb}	
\usepackage{caption}
\usepackage{longtable}
\usepackage{longfigure}

\title[ALMACAL VII: Interferometric Number Counts at 650$\mu$m]{ALMACAL VII: First Interferometric Number Counts at 650 $\mu$m}

\author[A. Klitsch et al.]{
A.~Klitsch$^{1, 2, 3}$\thanks{E-mail: anne.klitsch@gmail.com},
M.~A.~Zwaan$^{1}$,
Ian~Smail $^{2}$,
C.~P\'eroux$^{1, 4}$,
A.~D.~Biggs$^{1}$,
\newauthor{
Chian-Chou~Chen$^{5, 1}$,
R.~J.~Ivison$^{1}$,
G.~Popping$^{1}$,
C.~Lagos$^{6, 7, 8}$,
M.~B\'ethermin$^{4}$,}
\newauthor{
A.~M.~Swinbank $^{2}$,
A.~Hamanowicz$^{1}$,
R.~Dutta$^{1, 2, 9}$}
\\
$^{1}$European Southern Observatory, Karl-Schwarzschild-Str. 2, 85748 Garching near Munich, Germany\\
$^{2}$Centre for Extragalactic Astronomy, Durham University, Department of Physics, South Road, Durham DH1 3LE, UK\\
$^{3}$DARK, Niels Bohr Institute, University of Copenhagen, Lyngbyvej 2, 2100 Copenhagen, Denmark \\
$^{4}$Aix Marseille Univ, CNRS, LAM, (Laboratoire d'Astrophysique de Marseille), UMR 7326, 13388, Marseille, France\\
$^{5}$ Academia Sinica Institute of Astronomy and Astrophysics (ASIAA), No. 1, Section 4, Roosevelt Road, Taipei 10617, Taiwan \\
$^{6}$International Centre for Radio Astronomy Research (ICRAR), M468, University of Western Australia, 35 Stirling Hwy, Crawley, WA 6009, Australia.\\
$^{7}$ARC Centre of Excellence for All Sky Astrophysics in 3 Dimensions (ASTRO 3D).\\
$^{8}$Cosmic Dawn Center (DAWN).\\
$^{9}$Institute for Computational Cosmology, Durham University, South Road, Durham, DH1 3LE, UK
}

\date{Accepted XXX. Received YYY; in original form ZZZ}

\pubyear{2020}

\begin{document}
\label{firstpage}
\pagerange{\pageref{firstpage}--\pageref{lastpage}}
\maketitle

\begin{abstract}
Measurements of the cosmic far-infrared background (CIB) indicate that emission from many extragalactic phenomena, including star formation and black hole accretion, in the Universe can be obscured by dust. Resolving the CIB to study the population of galaxies in which this activity takes place is a major goal of submillimetre astronomy. Here, we present interferometric 650$\mu$m submillimetre number counts. Using the Band 8 data from the ALMACAL survey, we have analysed 81 ALMA calibrator fields together covering a total area of 5.5~arcmin$^2$. The typical central rms in these fields is $\sim 100 \mu$Jy~beam$^{-1}$ with the deepest maps reaching $\sigma = 47 \mu$Jy~beam$^{-1}$ at sub-arcsec resolution. Multi-wavelength coverage from ALMACAL allows us to exclude contamination from jets associated with the calibrators. However, residual contamination by jets and lensing remain a possibility. Using a signal-to-noise threshold of $4.5\sigma$, we find 21 dusty, star-forming galaxies with 650$\mu$m flux densities of $\geq 0.7 $mJy. At the detection limit we resolve $\simeq 100$ per cent of the CIB at 650$\mu$m, a significant improvement compared to low resolution studies at similar wavelength. We have therefore identified all the sources contributing to the EBL at 650 microns and predict that the contribution from objects with flux 0.7<mJy will be small.
\end{abstract}




\section{Introduction}

Twenty years ago, \citet{Puget1996} and \citet{Fixsen1998} measured the cosmic far-infrared background (CIB) with the {\it Cosmic Background Explorer} ({\it COBE}) Far Infrared Absolute Spectrometer (FIRAS), indicating that around half of the star-formation activity in the Universe is obscured by dust (see for example \citealt{Dole2006} and \citealt{Cooray2016} for a review and references therein for updated measurements). At the same time the field of galaxy formation and evolution was revolutionized by the discovery of submillimetre galaxies, a population of dusty star-forming galaxies with submillimetre flux densities of a few mJy which evolve strongly out to the high redshift Universe \citep[e.g.][]{Smail1997, Barger1998, Hughes1998, Ivison1998}. Resolving the CIB to study the population of galaxies in which this star formation takes place is a major research goal in submillimetre astronomy.

The abundance of galaxies above a certain flux threshold, the so called cumulative number counts ($N(>S)[{\rm deg^{-2}}]$), is a fundamental observable required to characterise galaxies and ultimately understand galaxy formation and evolution.
A challenge in measuring reliable number counts has for a long time been the large beam sizes ($\sim 15\arcsec - 30\arcsec$) for single-dish observations at far-infrared and submillimetre wavelengths. This low-spatial resolution results in bright confusion limits and can lead to source blending (often referred to as \textit{confusion}). Despite the number counts being a challenging quantity to measure, they can be used to shed light on galaxy formation \citep[e.g.][]{Baugh2005, Somerville2012, Lacey2016, Lagos2019}.


Major efforts have been undertaken to measure number counts at different far-infrared and sub/millimetre wavelengths. It is important to highlight that moving to shorter wavelengths provides measurements closer to the peak of the far-infrared spectral energy distribution (FIR SED) and therefore follows more closely the obscured star formation. At 1.1 and 1.2 mm, number counts have been determined using Bolocam at the Caltech Submillimeter Observatory (CSO) \citep{Laurent2005}, AzTEC on the James Clerk Maxwell Telescope (JCMT) and on the Atacama Submillimeter Telescope Experiment (ASTE) \citep[e.g.][]{Scott2010, Hatsukade2011, Scott2012, Umehata2014}. The Atacama Large (sub)Millimeter Array (ALMA) has also been used at this wavelength \citep[e.g.][]{Aravena2016a, Oteo2016, Umehata2017, Hatsukade2018, Franco2018, Simpson2020}. Submillimetre number counts at $850 \mu$m have been derived from surveys carried out with SCUBA \citep[e.g.][]{Blain1999, Chapman2002, Coppin2006} and later SCUBA-2 bolometer camera \citep[e.g.][]{Casey2013, Geach2017, Simpson2019} on the JCMT, the Large APEX Bolometer Camera (LABOCA) on the Atacama Pathfinder Experiment (APEX) \mbox{\citep{Beelen2008, Weiss2009}}. 
Number counts at $450 \mu$m and $500 \mu$m have been derived based on surveys using SCUBA, SCUBA-2 and \textit{Herschel}. Surveys aiming to constrain number counts at 450$\mu$m have either used gravitational lensing to magnify faint galaxies \mbox{\citep[e.g.][]{Smail2002, Chen2013}}, an untargeted-source extraction above the confusion limit \citep[e.g.][]{Oliver2010, Geach2013, Casey2013, Valiante2016, Wang2017}, stacking to constrain the faint end of the number counts \citep{Bethermin2012a} or ancillary data to construct a de-blended source catalogue \citep{Wang2019c}. Among the shortest wavelength studies, the number counts are not consistent between earlier and more recent results from \textit{Herschel}, perhaps indicating that some \textit{Herschel} source catalogues might be incorrectly de-blended.

Sophisticated techniques have been developed to characterize source blending. One approach which provided some insight into blending was to perform a higher resolution follow-up survey using radio interferometric observations to detect and resolve the counterparts \citep[e.g.][]{Ivison2002, Chapman2003, Chapman2005, Ivison2007}. More recently, such studies were extended to submillimetre interferometric follow-up of single-dish surveys to precisely identify counterparts and remove the effect of blending to construct more reliable source counts \citep[e.g.][]{Younger2009, Karim2013, Simpson2015, Simpson2020, Hill2018, Stach2018}. However, the multiplicity (the fraction of single-dish detections breaking up into multiple components at higher resolution) varies between $15$ and $>90$~per cent depending on factors such as source flux, survey depth, and definition of multiplicity. Therefore, number counts from single-dish observations are not yet robust.


To extend the number counts to fainter flux limits than the ALMA follow-up of single-dish surveys, to derive sub-mJy counts and assess the contribution to the extragalactic background light (EBL) we need interferometric surveys of blank fields. 
Such dedicated observing programmes were carried out with ALMA targeting cosmological deep fields to derive number counts at 1.1 and 1.2mm  such as ASPECS \citep{Aravena2016a, Aravena2019}, ALMA observations of the HUDF \citep{Dunlop2017}, the ASAGO survey in the GOODS-S field \citep{Hatsukade2018} and the GOODS-ALMA survey \citep{Franco2018}. All these targeted surveys are potentially subject to significant cosmic variance effects due to their small survey areas.

These studies are complemented by \citet{Oteo2016} who presented ALMA Band 6 (1.2mm) and Band 7 (850$\mu$m) number counts free of cosmic variance using archival ALMA calibration observations from the ALMACAL survey\footnote{\url{https://almacal.wordpress.com/}}. The authors used the 69 fields available at that time and applied a conservative source detection threshold. The authors were able to resolve 50~per cent of the EBL at these wavelengths.

The next step is to extend this approach to shorter wavelength to cover the peak of the EBL. The challenge will be to overcome low survey speeds of submillimetre interferometers because of their small primary beam. For example the full width at half maximum (FWHM) of the primary beam of ALMA at $650 \mu$m is only $12\arcsec$.

Here we expand on the work of \citet{Oteo2016} and perform an untargeted survey at $650 \mu$m free of source blending and cosmic variance. 
This work is bridging the gap between high resolution number counts at longer wavelengths and low resolution number counts available at shorter  wavelengths from \textit{Herschel} observations \citep[e.g.][]{Oliver2012, Valiante2016} and JCMT \citep[e.g.][]{Smail2002, Casey2013, Chen2013a, Chen2013, Hsu2016, Zavala2017, Wang2017}.

This paper is organized as follows: In \S 2 we describe the ALMACAL data reduction. Details of the source detection technique, the completeness, flux deboosting as well as the reliability of our sample of DSFGs are given in \S 3. In \S 4 we derive the number counts and calculate the contribution to the EBL. Finally, in \S 5 we summarize the main conclusion from this work.

\begin{figure*}
\centering{\textbf{Dusty star-forming galaxies}}\\
\vspace{0.2cm}
\includegraphics[width= 0.16\linewidth]{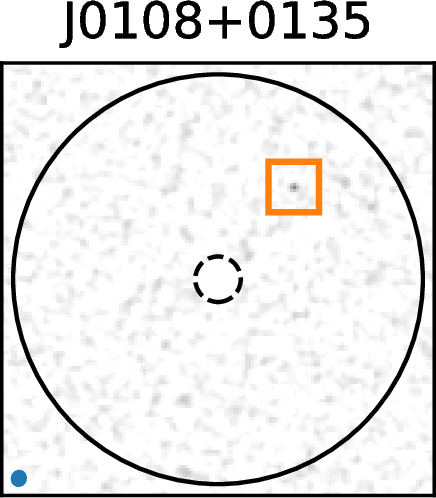}
\includegraphics[width= 0.16\linewidth]{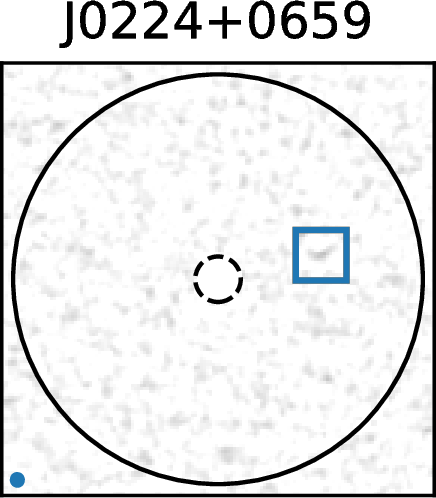}
\includegraphics[width= 0.16\linewidth]{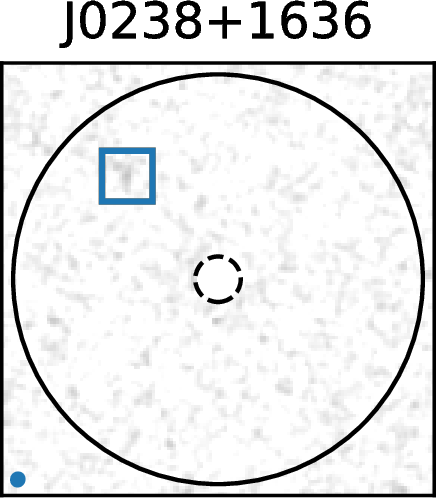}
\includegraphics[width= 0.16\linewidth]{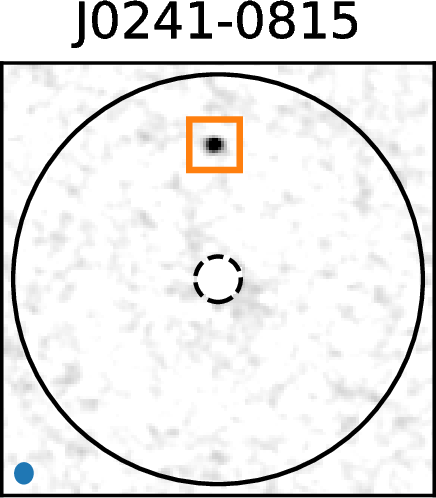}
\includegraphics[width= 0.16\linewidth]{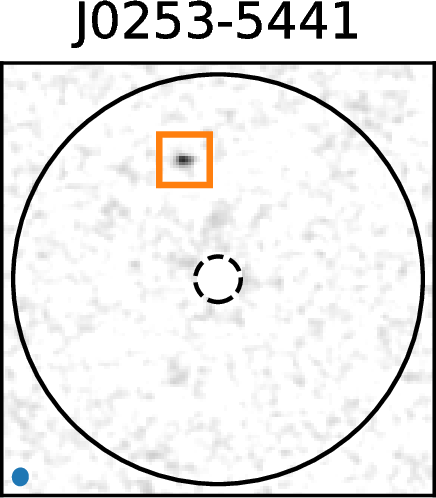}\\ \vspace{0.2cm}
\includegraphics[width= 0.16\linewidth]{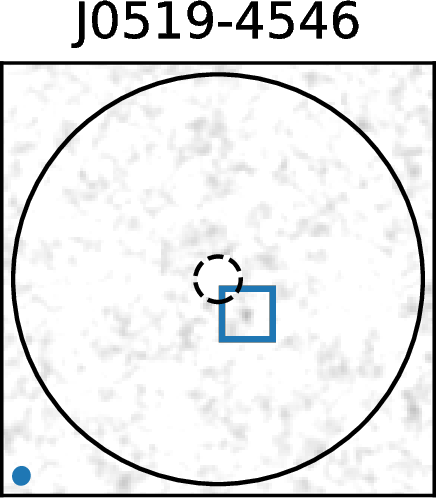}
\includegraphics[width= 0.16\linewidth]{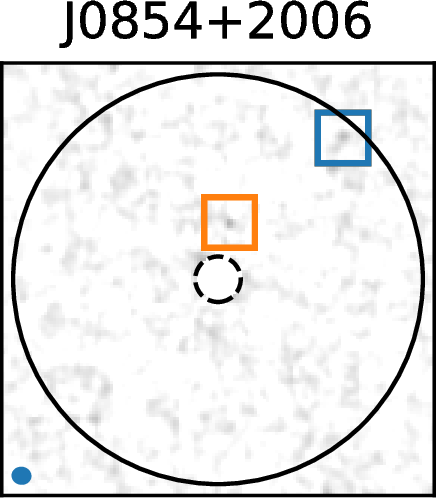}
\includegraphics[width= 0.16\linewidth]{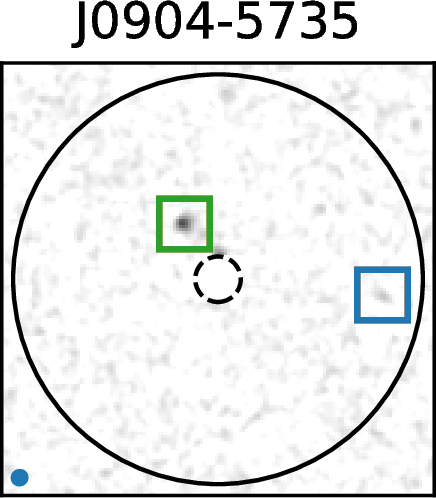}
\includegraphics[width= 0.16\linewidth]{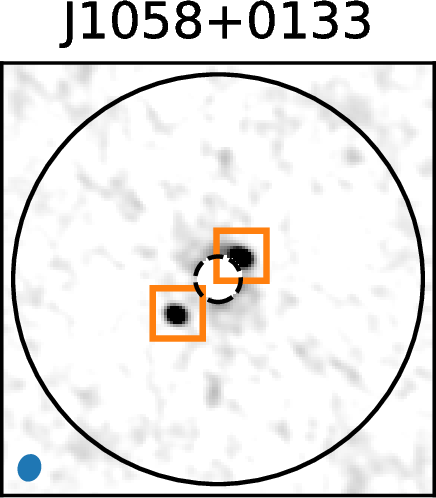}
\includegraphics[width= 0.16\linewidth]{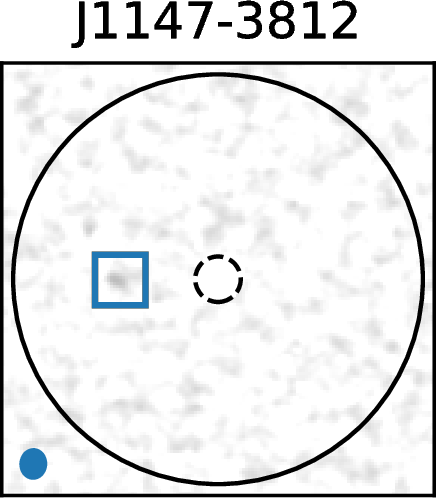}\\ \vspace{0.2cm}
\includegraphics[width= 0.16\linewidth]{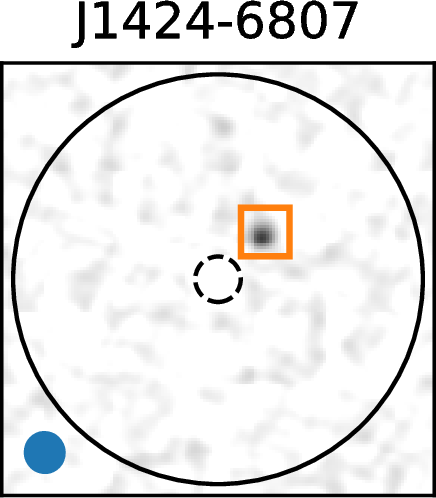}
\includegraphics[width= 0.16\linewidth]{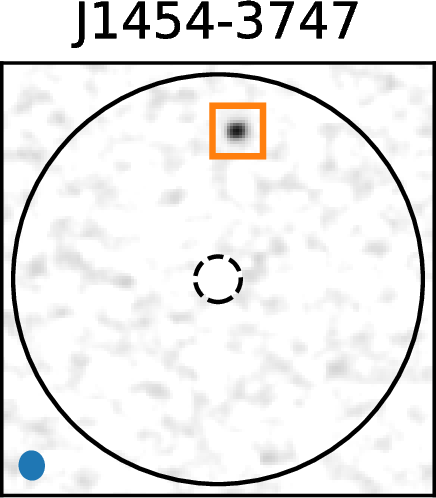}
\includegraphics[width= 0.16\linewidth]{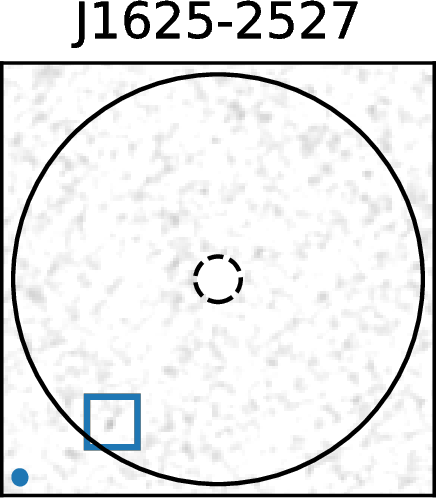}
\includegraphics[width= 0.16\linewidth]{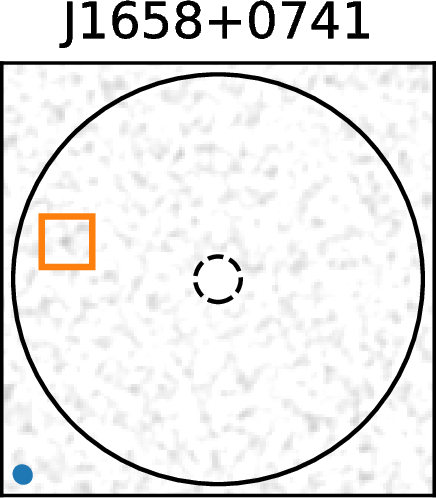}
\includegraphics[width= 0.16\linewidth]{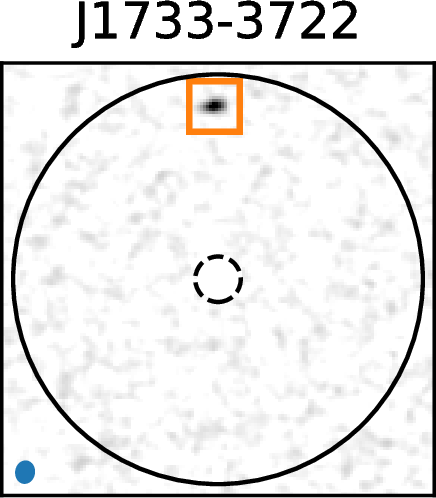}\\ \vspace{0.2cm}
\includegraphics[width= 0.16\linewidth]{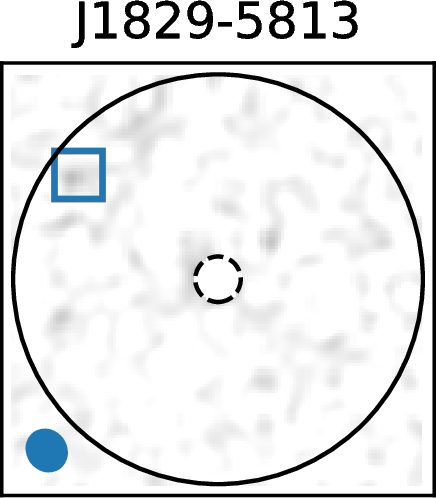}
\includegraphics[width= 0.16\linewidth]{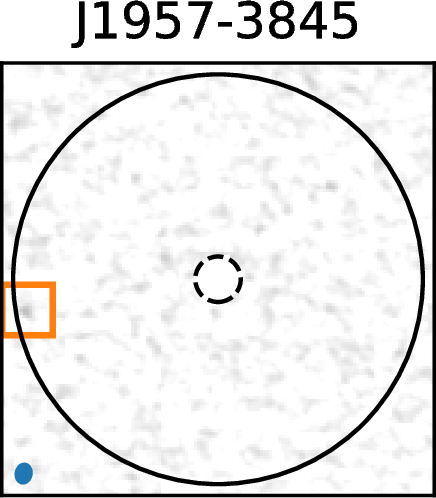}
\includegraphics[width= 0.16\linewidth]{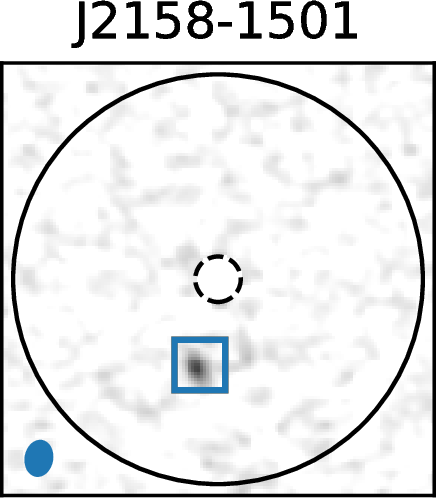}
\includegraphics[width= 0.16\linewidth]{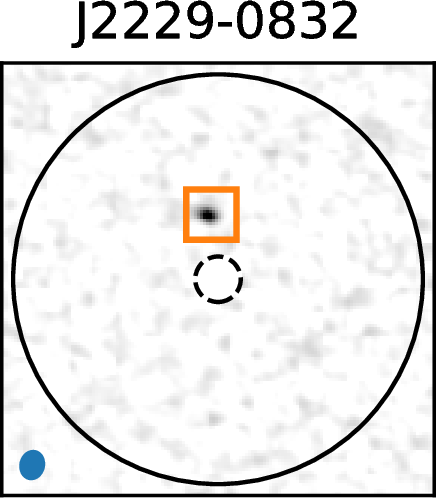} 
\includegraphics[width= 0.15\linewidth]{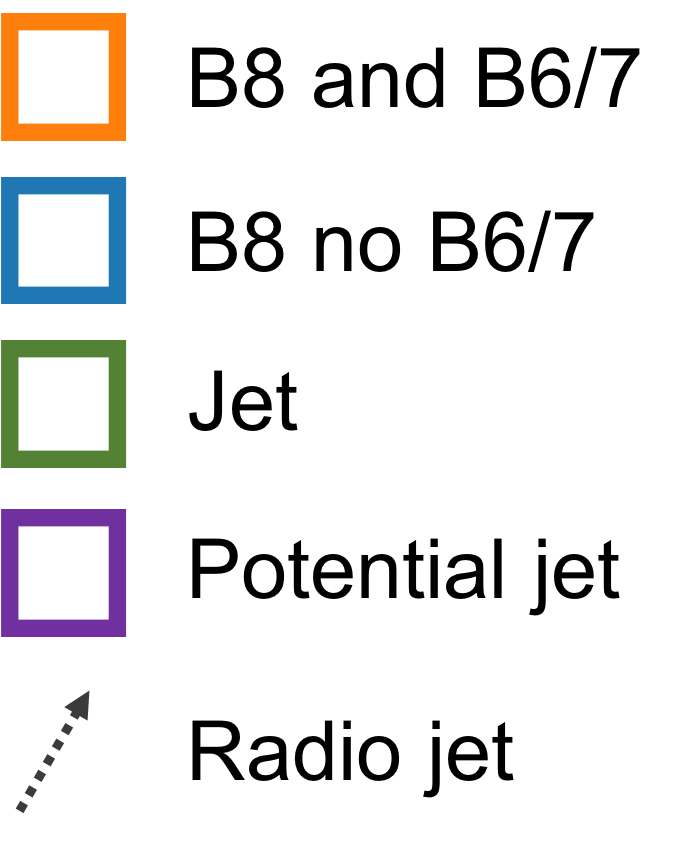}\\
\vspace{0.2cm}
\centering{\textbf{Jets}}\\
\vspace{0.2cm}
\includegraphics[width= 0.16\linewidth]{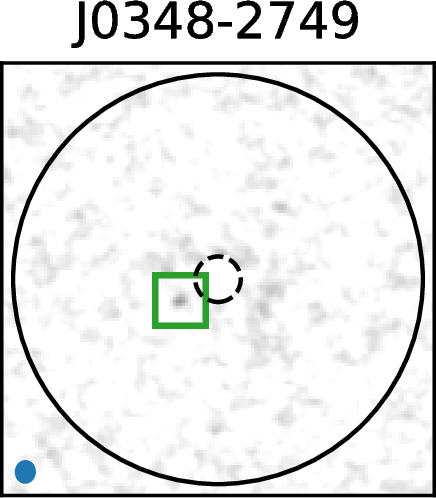}
\includegraphics[width= 0.16\linewidth]{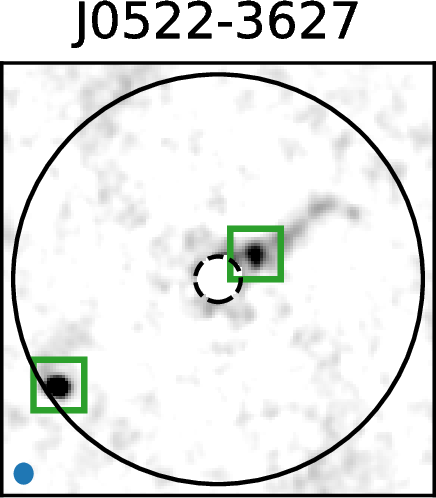}
\includegraphics[width= 0.16\linewidth]{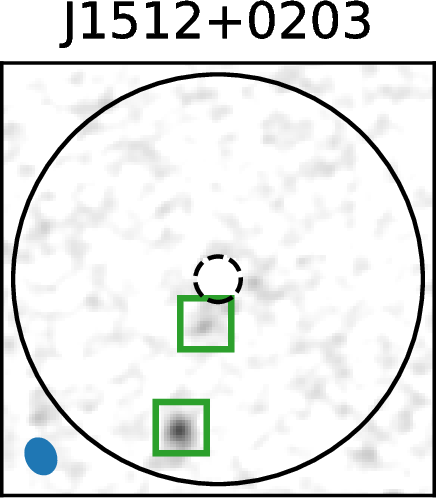}
\includegraphics[width= 0.16\linewidth]{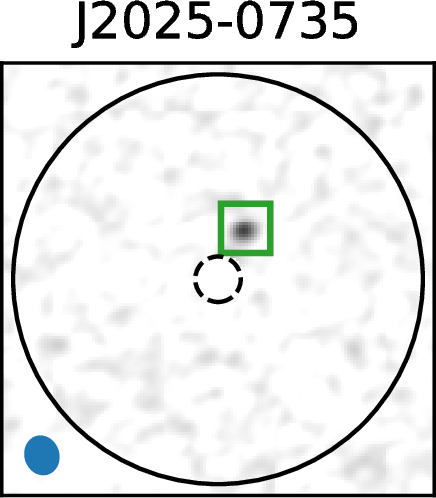}
\includegraphics[width= 0.16\linewidth]{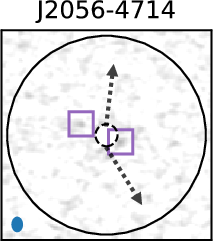}\\ \vspace{0.2cm}
\includegraphics[width= 0.16\linewidth]{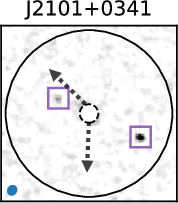}
\includegraphics[width= 0.16\linewidth]{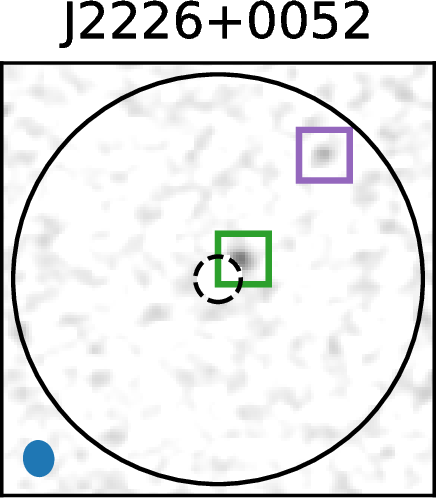}
\includegraphics[width= 0.16\linewidth]{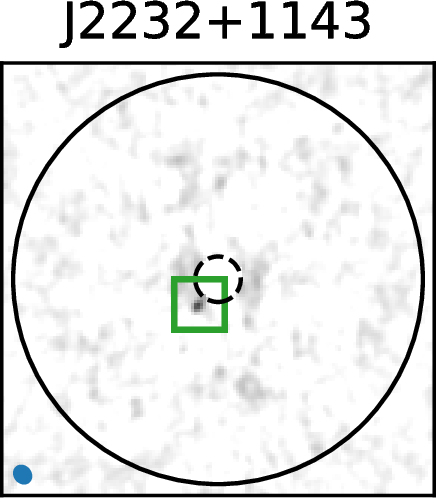}
\includegraphics[width= 0.16\linewidth]{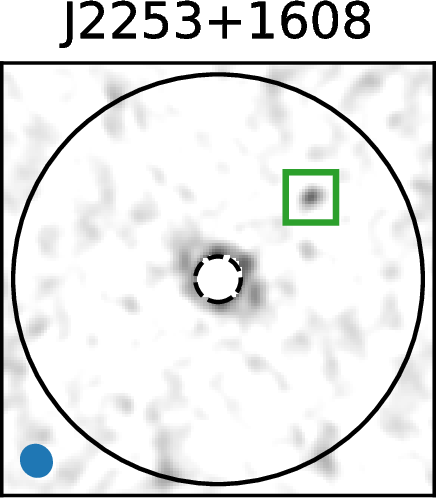}
\includegraphics[width= 0.16\linewidth]{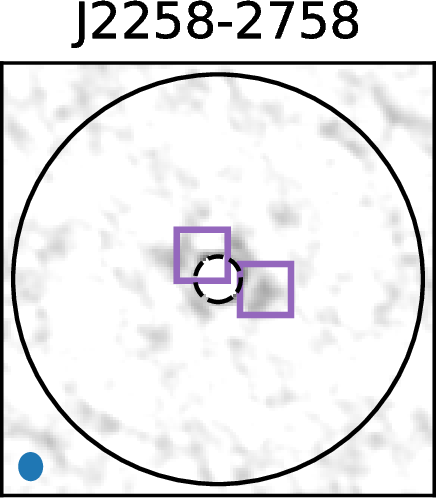}

\caption[Band 8 maps of the $0.3\arcsec$ resolution detections.]{650$\mu$m (ALMA Band 8) signal-to-noise maps of the detections in the $0.3\arcsec$ resolution images. Band 8 detections are marked with squares (Band 8 and Band 6 detection: orange, Band 8 with upper limit in Band 6 and/or Band 7 detection: blue, identified jets: black, potential jets (see text): grey), the solid black circle indicates the area over which we search for continuum emission ($1.5$ times the primary beam FWHM in Band 8) corresponding to $18\arcsec$ diameter, the dashed black circle indicates the central $2\arcsec$ diameter region excluded from the survey due to possible contamination from quasar residuals. Black dashed arrows mark the direction of jets identified from archival VLA published ATCA maps. The blue ellipse illustrates the synthesized beam. \label{Chap6FigDetections}}
\end{figure*}

\section{ALMACAL Data Reduction}

We analyse all ALMA calibrator observations prior to 2018 December. The observed fields are distributed quasi-randomly on the sky visible from the Atacama desert. All ALMA calibration data using Band 8 (385 -- 500 GHz) are included in our dataset comprising a total of 112h of observing time. Therefore, the only biases we introduce are due to the observing latitude of ALMA, the annual weather patterns in the Atacama desert and the positions of sources of interest for studies by the astronomical community in Band 8, e.g.\ the cosmological deep fields. The data retrieval is described in detail in \citet{Oteo2016}. Since we are interested in high resolution number counts we select only observations with a spatial resolution $< 1\arcsec$. Furthermore, we use only fields for which we reach a central rms of $\lesssim 1 {\rm mJy \; beam^{-1}}$ in the combined maps of the same fields to avoid images with lower data quality. For this analysis we use pseudo-continuum visibilities which are already integrated over the spectral dimension and therefore require less storage space and allow faster processing.

The calibrator-subtracted pseudo-continuum visibilities from every execution block are first imaged individually without combining data for a given calibrator. We visually inspect every map and discard those showing signs of poor calibration (e.g.\ stripes or significant halos and residuals around the quasar position). The drawback of using pseudo-continuum visibilities is that the weights of the visibilities cannot be recalculated. To ensure an equal representation of all observations  in the combined image it is necessary that all weights are on the same scale.
Therefore, we also inspect the weights of the visibilities and include only those observations with weights similar to the average weights for a given calibrator. This leads to the loss of $\sim 26$~per cent of the data but allows a homogeneous treatment. Due to the overall dataset size and complexity, flagging and recalibration are impracticable.

The data reduction is carried out using the {\sc Common Astronomy Software Application ({\sc CASA})} \citep{McMullin2007} version 5.5.0. We combine the data for each calibrator using the task {\sc concat} and image the combined visibilities using the task {\sc tclean}. We define cleaning windows using the automatic masking procedure ``auto-multithresh''. 
A natural weighting scheme is chosen to ensure optimal use of all baselines, resulting in the lowest possible r.m.s. noise. To avoid resolving the galaxies we set the outer taper to $0.3\arcsec$ similar to the scale of dust emission in DSFGs \citep{Simpson2015, Gullberg2019}. We produce a second set of images with an outer taper of $0.8\arcsec$ to test if we are missing detections because they are resolved out in the higher resolution imaging and to confirm the reliability of source fluxes from the higher resolution maps. The maps without primary beam correction are used for the source detection and subsequent statistical analysis. However, for the final flux measurements, we correct for the primary beam attenuation using the task {\sc impbcor}.

We image ALMACAL observations of 81 calibrators observed in Band 8 covering a total of $\sim 5.5$~arcmin$^2$ within $1.5$ times the FWHM of the primary beam. Examples of these are shown in Fig.~\ref{Chap6FigDetections}. We note that bandwidth smearing is negligible at this radial distance. The effective survey area is a function of the source flux density. 
We reach noise levels of  $47 - 1022\mu$Jy~beam$^{-1}$ with a median of $187\mu$Jy~beam$^{-1}$ and resolutions of $0.34\arcsec - 0.98\arcsec$ with a median of $0.53\arcsec$. The r.m.s. in the ALMACAL Band 8 maps is significantly higher than that in Band 6 and 7 because of higher receiver and sky noise and typically shorter exposure times. The mean wavelength of all ALMACAL observations in Band 8 used here is 650$\mu$m.

\section{Analysis}

\subsection{Source Detection}
\label{Cahp6SecSourceDetection}

In Fig.~\ref{Chap6FigDetections} we show the ALMA Band 8 maps in which we detect continuum sources using the following procedure. 
We perform the source detection using SExtractor \citep{Bertin1996} on the clean maps before correcting for the primary beam attenuation, to ensure uniform noise properties. The calibrators are seen as bright sources in the centres of the Band 8 maps. We detect residual signal in Band 8 from the calibrators more frequently than in the Band 6 and Band 7 maps presented by \citet{Oteo2016}. We mask the central region of each map with a radius of $1 \arcsec$ and exclude this region from further analysis. The radius is chosen based on visual inspection of the maps. Furthermore, we use a detection threshold with a peak flux of at least $4.5$ times the r.m.s. noise in the image, comparable with those used in previous studies \citep{Simpson2015, Oteo2016, Stach2019}. At this threshold we are able to detect galaxies down to $\sim0.7$~mJy and at the same time keep the number of spurious detections to a minimum.

\begin{figure*}
\center
\includegraphics[width = 0.325\linewidth]{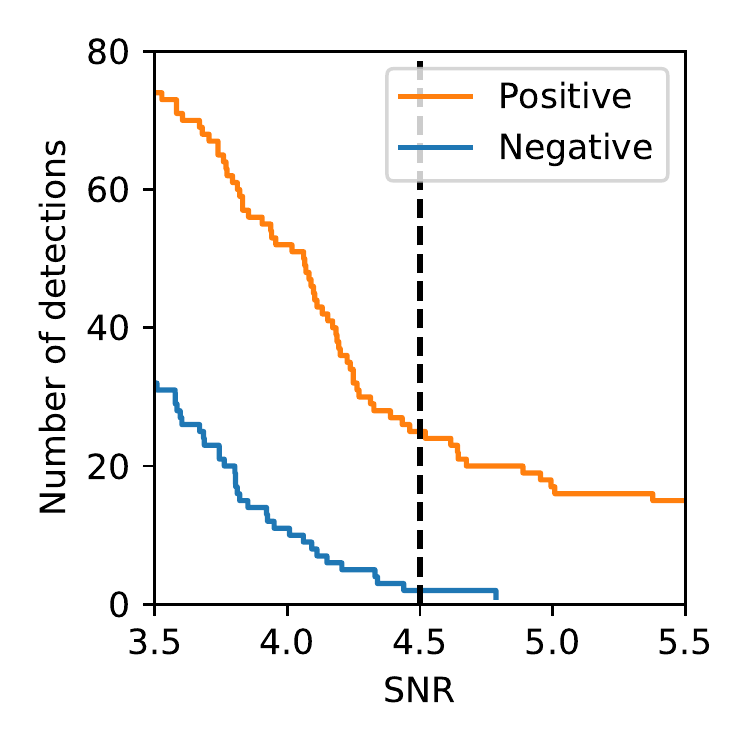}
\includegraphics[width = 0.325\linewidth]{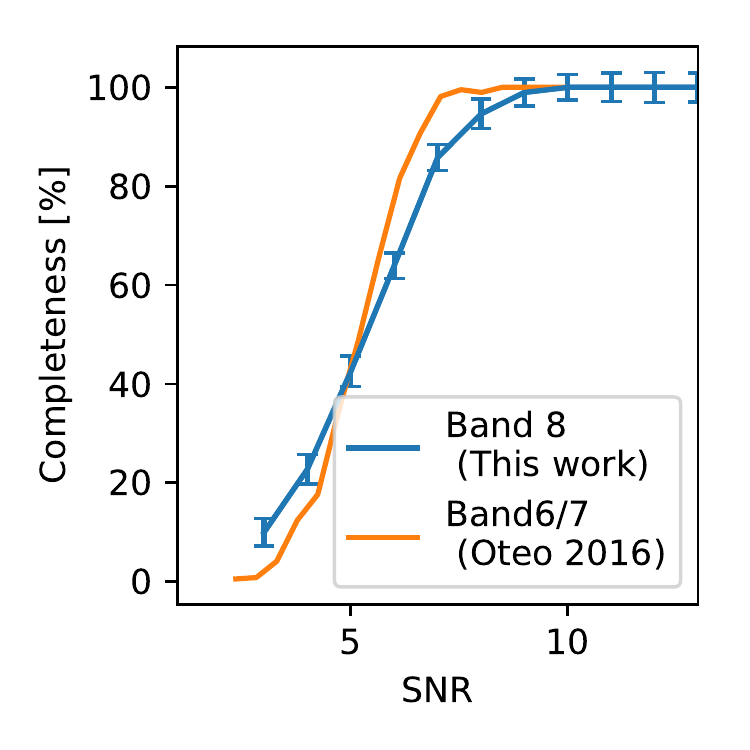}
\includegraphics[width = 0.325\linewidth]{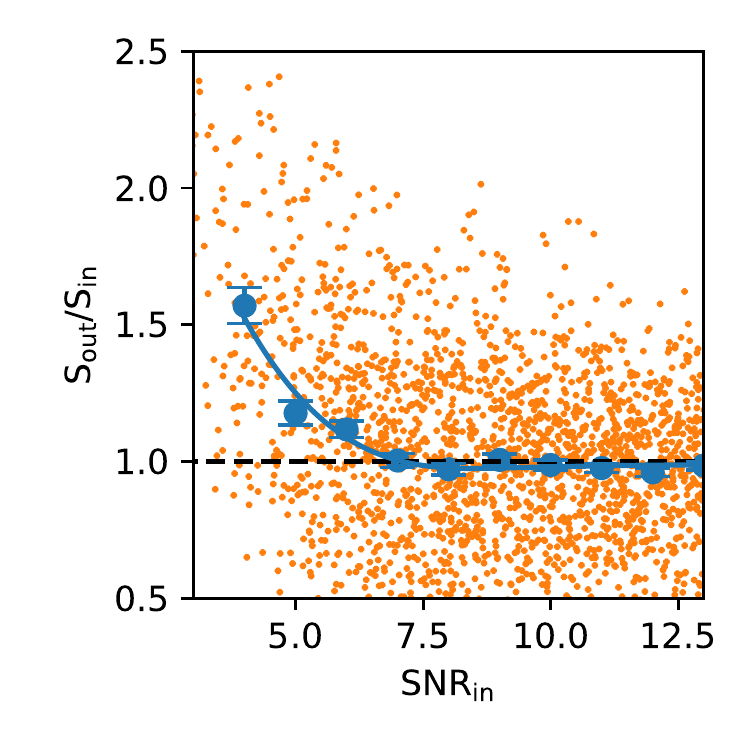}
\caption[The cumulative number of spurious detections in the inverted maps as a function of the pear SNR.]{\textit{Right panel:} The cumulative number of spurious detections in the inverted maps as a function of the peak SNR\@. The highest significance spurious detection is at $4.7\sigma$. Therefore, we choose a detection threshold of $4.5 \sigma$ (black dashed line) corresponding to expecting one false detection. Combined with the multi-band detection this offers a high reliability of our detections. \textit{Middle panel:} Completeness of ALMACAL Band 8 as a function of the SNR of the detected sources. We reach a completeness of 100~per cent at $9 \sigma$ and a completeness of 80~per cent at a round $7\sigma$. \textit{Left panel:} The ratio between the output and input flux densities of simulated sources as a function of the input SNR (defined as the ratio between the input flux density and the r.m.s. at the centre of the map). The output flux densities tend to be increasingly overestimated at a SNR $< 7 \sigma$. At $4.5\sigma$ the flux boosting is 38~per cent and we correct our measured fluxes for this.\label{Chap6FigFluxBoosting}  \label{Chap6FigReliability} \label{Chap6FigCompleteness}}
\end{figure*}

Performing source detection using a modest detection threshold leads to the detection of spurious noise peaks. Since we are aiming for a reliable number counts measurement, we choose to include only high reliability detections. To test the reliability as a function of SNR we invert our maps and run the source finder with the same parameters as for the main search. Any detection in the inverted map is considered to be a spurious noise peak. The cumulative distribution of noise peaks as a function of SNR is shown in Fig.~\ref{Chap6FigReliability}. We find that the highest SNR detection in the inverted map is at $ 4.7 \sigma$. Therefore, at our detection threshold of $4.5 \sigma$ the contamination from spurious sources is negligible. 

%

For each Band 8 detection we make maps from the ALMACAL Band 6 and 7 observations to confirm detections via multi-band observations and measure the slope of the spectral energy distribution (SED)\@. The FWHM of the primary beam in Band 6 and 7 is wider than in Band 8 (FWHM$_{\rm B6} =  27\arcsec$, FWHM$_{\rm B7} =  18\arcsec$, FWHM$_{\rm B8} =  12\arcsec$). Therefore, any detection in Band 8 will be covered by the Band 6 and 7 observations if the calibrator has been observed at both wavelengths and the Band 6 and/or 7 data are sufficiently deep.

\subsection{Effective Area}\label{Chap6SecArea}

\begin{figure}
    \centering
    \includegraphics[width = \linewidth]{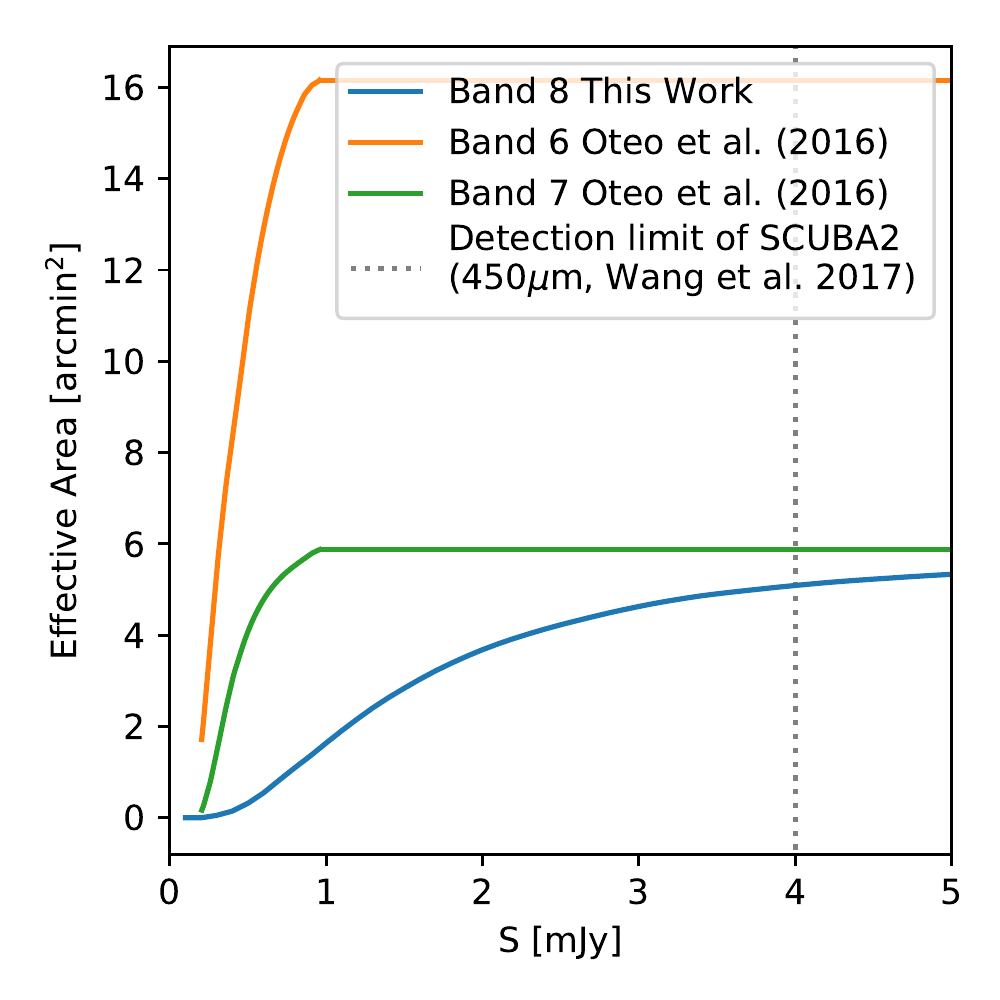}
    \caption[The effective area covered by the current ALMACAL Band 8 observations as a function of the flux density of detected sources.]{The effective area covered by the current ALMACAL Band 8 observations as a function of $4.5 \sigma$ flux limit for detecting sources. We calculate the area over which a galaxy could be detected at a peak signal-to-noise ratio (SNR) of $4.5\sigma$ in each map and sum these. For comparison, we show the effective area probed in the Band 6 and Band 7 study by \citet{Oteo2016} which benefits from the larger primary beam in Band 6 and 7. Only with the multi field observations offered by ALMACAL can we derive robust number counts at short wavelengths.\label{Chap6FigEffectiveArea}}
\end{figure}

The sensitivity in an interferometric observation (such as our ALMA observations) is not uniform within the field of view, but decreases with increasing distance from the centre due to the primary beam response. The effective area over which a galaxy can be detected is therefore a function of the flux density. We define the maximum extent of a map to be $1.5$~times the FWHM ($12\arcsec$) of the primary beam expected at Band 8.

We measure the effective area as a function of SNR for our survey (shown in Fig.~\ref{Chap6FigEffectiveArea}). Here we exclude the central region that is potentially contaminated by residuals from the calibrators. We reach an effective area of $\sim 5.5$~arcmin$^2$ for a flux density of $\geqslant 5$~mJy ($4.5 \sigma$). In the earlier work using ALMACAL \citet{Oteo2016} achieved effective areas of $\sim 6$ and $\sim 16$~arcmin$^2$ for a flux density of $1$~mJy at Band 7 and Band 6, respectively. That study benefited from,  on average, deeper maps in Band 6 and 7 due to lower receiver and sky noise as well as the much wider field of view at longer wavelengths (FoV $\propto \lambda^2$).

This work using multi-field observations from ALMACAL is a unique opportunity to derive number counts at shorter wavelengths from high resolution observations, free of source blending and cosmic variance.

\subsection{Assessing Sample Completeness}\label{Chap6SecCompleteness}


We use artificial sources to measure the completeness of our survey as a function of the signal-to-noise ratio (SNR)\@. We inject artificial point sources with uniformly distributed random fluxes between 2 and 25~times the noise level in the ALMACAL visibility data. The sources are put in at random positions within the 1.5~times~FWHM of the primary beam search radius. We inject 20 sources per map and repeat this procedure 50 times. 
To test the full data reduction and analysis chain, the visibilities with the injected artificial sources are imaged using the same settings for the {\sc casa} task {\sc tclean} as for the original visibilities. We then use the same source finding procedure as for the real data to recover the artificial sources. In case a source was injected within a radius of six times the beam width from another artificial source or within a radius of $1\arcsec$ from the centre it is excluded from further analysis. A source is considered to be recovered if it is detected with SExtractor at $\geq 2.5 \sigma$ and within one synthesized beam width from the position of the injection. To estimate the errors on the completeness, we perform a bootstrap resampling. We take the parent population of $n$ artificial sources and replace those with $n$ randomly selected sources. This process is repeated 200 times and the completeness is calculated for each realization. We determine the scatter of the different realizations of the completeness. The resulting completeness as a function of SNR is shown in Fig.~\ref{Chap6FigCompleteness}. 

Our survey is 100~per cent complete at a SNR $\geq 9$ and 50~per cent complete at an SNR $\geq 5$. Compared to the dusty star-forming galaxy (DSFG) survey in Band 6 and 7 presented by \citet{Oteo2016} our completeness function is slightly flatter reaching a high completeness at higher SNR\@. This is due to the fact that the details of the analysis were chosen in a slightly different way. Furthermore, we have shorter observing times per field in Band 8 than in Band 6 and 7. At shorter wavelengths, the $uv$ coverage is not as good as at longer wavelengths and the noise in the ALMA maps becomes more non-Gaussian than at longer wavelengths.

We assess the possibility of missing detections due to the high resolution in Section~\ref{Chap6SecSourceCat}.

\subsection{Flux Deboosting}


A known issue of measuring flux densities of continuum sources detected at low SNR is the fact that their flux densities can be boosted due to the presence of noise fluctuations \citep[e.g.][]{Coppin2006}. To measure this effect we use the same set of artificial point sources described in Section~\ref{Chap6SecCompleteness}. We measure the flux density of the detected sources relative to the input flux density prior to primary beam correction. 
The flux measurement is performed in the same way as for the real detections. 
The results are shown in Fig.~\ref{Chap6FigFluxBoosting}. We find that the flux density for sources detected at $4.5 \sigma$ is boosted on average by $\sim 38$~per cent. At a SNR of $\geq 7$ the effect of flux boosting is negligible. 
Half of our DSFG detections fall in the regime below $7 \sigma$ where flux boosting needs to be corrected. We resample the measured flux boosting in bins of 0.25 and fit a cubic spline to the mean in each bin. We correct the flux densities of our catalogue based on the value of the fitted spline at the SNR of each source.

\subsection{Source Catalogue}\label{Chap6SecSourceCat}

\begin{table*}

\caption[Physical properties of DSFGs detected at $650\mu$m up to December 2018 in our ALMACAL survey.]{Properties of 38 continuum detections (21 DSFGs and seven potential jets and ten jets) detected at $650\mu$m and calibrators in our ALMACAL survey up to December 2018.\label{Chap6TabDetections}}
\begin{tabular}{l c c c c c c c c}

    \hline
      Name & $z_{\rm cal}$ & $S_{650 \mu {\rm m}}$ & $\theta_{\rm B8}$ & $S_{870 \mu{\rm m}}$ & $S_{650 \mu {\rm m}} / S_{870 \mu {\rm m}}$ & $S_{1.2 {\rm mm}}$ & $S_{650 \mu {\rm m}} / S_{1.2 {\rm mm}}$ \\
       & & [mJy] & [$\arcsec \times \arcsec$] & [mJy] & & [mJy] & \\
      \hline
      
\textbf{DSFG}\\
ALMACAL 010838.56+013504.3 & 2.099    & 3.6 $\pm$ 0.6  	& 0.38 $\times$ 0.35	& 3.2 $\pm$ 0.1		& 1.2 $\pm$ 0.2	 & 0.6 $\pm$ 0.2	& 5.7 $\pm$ 2.2	    	\\ 
ALMACAL 022428.13+065924.3 & 0.511    & 1.3 $\pm$ 0.3  	& 0.34 $\times$ 0.33	& ... 	  		& ... 	    	 & $<$0.7  		& $>$1.9   	      	\\ 
ALMACAL 023839.21+163703.7 & 0.94     & 2.0 $\pm$ 0.2  	& 0.35 $\times$ 0.34	& 0.3 $\pm$ 0.1 	& 7.4 $\pm$ 3.8	 & $<$0.3		& $>6.7$	    	\\ 
ALMACAL 024104.81-081514.9 & 0.00512  & 3.0 $\pm$ 0.1  	& 0.49 $\times$ 0.43	& 2.0 $\pm$ 0.1 	& 1.5 $\pm$ 0.1	 & 0.5 $\pm$ 0.1 	& 5.6 $\pm$ 1.0	    	\\ 
ALMACAL 025329.36-544146.3 & 0.539    & 2.1 $\pm$ 0.1  	& 0.41 $\times$ 0.37	& 1.5 $\pm$ 0.1 	& 1.4 $\pm$ 0.1	 & 0.3 $\pm$ 0.1 	& 6.4 $\pm$ 3.2	    	\\ 
ALMACAL 051949.61-454645.5 & 0.035    & 0.5 $\pm$ 0.2  	& 0.43 $\times$ 0.40	& $<$0.2    		& $>$2.5     	 & $<$0.1    		& $>5$ 	    	      	\\ 
ALMACAL 085448.49+200636.8 & 0.306    & 5.0 $\pm$ 0.5  	& 0.43 $\times$ 0.39	& $<$0.4 		& $>$12.1     	 & $<$0.4     		& $>12.5$    	       	\\ 
ALMACAL 085448.85+200633.0 & 0.306    & 2.0 $\pm$ 0.6  	& 0.43 $\times$ 0.39	& 0.3 $\pm$ 0.1 	& 7.0 $\pm$ 3.4	 & 0.7 $\pm$ 0.1 	& 2.8 $\pm$ 1.0	    	\\ 
ALMACAL 090452.29-573506.6 & 0.695    & 4.2 $\pm$ 0.9  	& 0.39 $\times$ 0.38	& $<$2.8    		& $>$1.5      	 & $<$5.2     		&$>0.8$     	       	\\ 
ALMACAL 105829.73+013357.2 & 0.888    & 7.2 $\pm$ 0.6   & 0.61 $\times$ 0.51	& 6.5 $\pm$ 0.2 	& 1.1 $\pm$ 0.1	 & 2.0 $\pm$ 0.1	& 3.6 $\pm$ 0.3	   	\\ 
ALMACAL 105829.54+013359.8 & 0.888    & 13.4 $\pm$ 0.6  & 0.61 $\times$ 0.51	& 4.4 $\pm$ 0.2		& 3.0 $\pm$ 0.2	 & 1.5 $\pm$ 0.1	& 8.9 $\pm$ 0.7	   	\\ 
ALMACAL 114701.74-381211.2 & 1.048    & 2.2 $\pm$ 0.6  	& 0.69 $\times$ 0.61	& $<$0.5    		& $>$4.7	 & $<$0.3        	& $>7.3$	   	\\ 
ALMACAL 142455.22-680756.2 & ...      & 3.8 $\pm$ 0.5  	& 0.94 $\times$ 0.92	& 2.3 $\pm$ 0.2 	& 1.7 $\pm$ 0.3	 & 1.1 $\pm$ 0.1 	& 3.6 $\pm$ 0.6	    	\\ 
ALMACAL 145427.34-374726.7 & 0.31421  & 9.4 $\pm$ 0.4  	& 0.66 $\times$ 0.57	& ...       		& ... 	    	 & 2.3 $\pm$ 0.1 	& 4.1 $\pm$ 0.3	        \\ 
ALMACAL 162547.24-252744.7 & 0.786    & 0.9 $\pm$ 0.3  	& 0.40 $\times$ 0.37	& 0.2 $\pm$ 0.1 	& 3.6 $\pm$ 2.3	 & $<$0.3    		& $>3$   	    	\\ 
ALMACAL 165809.46+074129.1 & 0.621    & 7.2 $\pm$ 1.6  	& 0.46 $\times$ 0.43	&   	    	  	&      	    	 & ...			& ... 	    	   	\\ 
ALMACAL 173315.21-372224.9 & ...      & 11.0 $\pm$ 0.2 	& 0.51 $\times$ 0.43	& 8.6 $\pm$ 0.2 	& 1.3 $\pm$ 0.0	 & 2.2 $\pm$ 0.1 	& 5.0 $\pm$ 0.3	    	\\ 
ALMACAL 182913.20-581350.8 & 1.531    & 2.5 $\pm$ 0.4  	& 0.99 $\times$ 0.89	& ... 	    	      	& ...      	 & $<$0.7    	 	& $>3.6$     	    	      	    \\ 
ALMACAL 195800.54-384507.8 & 0.63     & 2.3 $\pm$ 0.2  	& 0.48 $\times$ 0.39	&...		  	& ...		 & ...			& ...		   	\\ 
ALMACAL 215806.35-150113.3 & 0.67183  & 2.9 $\pm$ 0.2  	& 0.82 $\times$ 0.62	& ... 	    	      	& ...		 & $<$1.2    	 	& $>2.4$     	    	\\ 
ALMACAL 222940.12-083251.8 & 1.5595   & 4.9 $\pm$ 0.4  	& 0.67 $\times$ 0.56	& 3.4 $\pm$ 0.1 	& 1.4 $\pm$ 0.1	 & 1.6 $\pm$ 0.2 	& 3.1 $\pm$ 0.4	    	\\ 
										                                      
\hline										                                      
\textbf{Potential Jets}\\							                                      
ALMACAL 205616.59-471446.7 & 1.489    & 1.0 $\pm$ 0.4  	& 0.71 $\times$ 0.53	& $<$0.4 &		 $>$2.7     	 & 0.6 $\pm$ 0.1	& 1.6 $\pm$ 0.7	      	 \\ 
ALMACAL 205616.24-471448.3 & 1.489    & 1.5 $\pm$ 0.4  	& 0.71 $\times$ 0.53	& 0.5 $\pm$ 0.1 	 & 2.8 $\pm$ 1.0 & 1.2 $\pm$ 0.1 	& 1.3 $\pm$ 0.4	   	 \\ 
ALMACAL 210139.06+034132.9 & 1.013    & 1.4 $\pm$ 0.2  	& 0.60 $\times$ 0.53	& 0.5 $\pm$ 0.2 	 & 2.5 $\pm$ 1.0 & 0.7 $\pm$ 0.1	& 2.1 $\pm$ 0.6	   	 \\ 
ALMACAL 210138.47+034128.7 & 1.013    & 5.8 $\pm$ 0.2  	& 0.60 $\times$ 0.53	& 2.1 $\pm$ 0.2 	 & 2.8 $\pm$ 0.3 & 0.8 $\pm$ 0.1 	& 7.5 $\pm$ 1.4	   	 \\ 
ALMACAL 222646.23+005216.7 & 2.25     & 0.6 $\pm$ 0.2  	& 0.81 $\times$ 0.69	& 0.9 $\pm$ 0.1 	 & 0.6 $\pm$ 0.2 & 0.3 $\pm$ 0.1 	& 1.8 $\pm$ 0.6	   	 \\ 
ALMACAL 225806.02-275820.3 & 0.92562  & 2.1 $\pm$ 0.2  	& 0.63 $\times$ 0.54	& $<$0.2   		 & $>$9.0     	 & 1.6 $\pm$ 0.1 	& 1.3 $\pm$ 0.1	      	 \\ 
ALMACAL 225805.81-275821.8 & 0.92562  & 1.1 $\pm$ 0.2  	& 0.63 $\times$ 0.54	& $<$0.2    	 	 & $>$4.5        & 1.9 $\pm$ 0.1 	& 0.6 $\pm$ 0.1	           \\

\hline										                                      
\textbf{Jets}\\									                                      
ALMACAL 034838.28-274914.6 & 0.176    & 0.7 $\pm$ 0.2  	& 0.52 $\times$ 0.46	& 1.3 $\pm$ 0.1		& 0.5 $\pm$ 0.1	 & 4.0 $\pm$ 0.2	& 0.2 $\pm$ 0.1	    	\\ 
ALMACAL 052257.86-362729.8 & 0.05629  & 8.7 $\pm$ 0.5  	& 0.48 $\times$ 0.44	& 7.1 $\pm$ 0.6 	& 1.2 $\pm$ 0.1	 & 24.8 $\pm$ 2.9 	& 0.3 $\pm$ 0.1     	\\ 
ALMACAL 052258.57-362735.6 & 0.05629  & 45.8 $\pm$ 0.4 	& 0.48 $\times$ 0.44	& 43.8 $\pm$ 0.5 	& 1.0 $\pm$ 0.1  & 61.0 $\pm$ 2.3 	& 0.8 $\pm$ 0.1    	\\ 
ALMACAL 090453.37-573503.4 & 0.695    & 16.1 $\pm$ 1.0 	& 0.39 $\times$ 0.38	& 24.1 $\pm$ 0.7 	& 0.7 $\pm$ 0.1  & 37.1 $\pm$ 1.4 	& 0.4 $\pm$ 0.1    	\\ 
ALMACAL 151215.86+020310.3 & 0.219    & 4.0 $\pm$ 0.2  	& 0.86 $\times$ 0.68	& 5.0 $\pm$ 0.2 	& 0.8 $\pm$ 0.1	 & 5.9 $\pm$ 0.2 	& 0.7 $\pm$ 0.0	    	\\ 
ALMACAL 151215.79+020314.9 & 0.219    & 0.7 $\pm$ 0.2  	& 0.86 $\times$ 0.68	& 0.5 $\pm$ 0.3 	& 1.4 $\pm$ 0.9	 & 0.8 $\pm$ 0.3 	& 0.9 $\pm$ 0.4	    	\\ 
ALMACAL 202540.59-073550.6 & 1.388    & 2.2 $\pm$ 0.3  	& 0.88 $\times$ 0.76	& 3.8 $\pm$ 0.3 	& 0.6 $\pm$ 0.1	 & 4.0 $\pm$ 0.1 	& 0.5 $\pm$ 0.1	    	\\ 
ALMACAL 222646.47+005212.1 & 2.25     & 1.3 $\pm$ 0.2  	& 0.81 $\times$ 0.69	& 0.3 $\pm$ 0.1 	& 4.0 $\pm$ 1.3	 & 0.4 $\pm$ 0.1 	& 3.5 $\pm$ 0.7	    	\\ 
										                                      
ALMACAL 223236.47+114349.7 & 1.037    & 2.2 $\pm$ 0.2  	& 0.46 $\times$ 0.39	& 2.4 $\pm$ 0.2 	& 0.9 $\pm$ 0.1	 & 4.8 $\pm$ 0.2 	& 0.5 $\pm$ 0.1	    	\\ 
ALMACAL 225357.47+160857.1 & 0.859    & 1.3 $\pm$ 0.4  	& 0.76 $\times$ 0.70	& 3.7 $\pm$ 0.4 	& 0.3 $\pm$ 0.1	 & 6.0 $\pm$ 0.3 	& 0.2 $\pm$ 0.1	    	\\ 

\end{tabular}\\
\flushleft
\textbf{Notes: } Sources in the upper part of the Table are identified DSFGs, sources in the middle part are potential jets based on their spatial alignment with the calibrator and sources in the lower part are jets based on their submillimetre colors. 
\end{table*}

In the 81 ALMACAL Band 8 maps, we found 38 continuum detections in Band 8. Four of the new Band 8 detections were already detected in a previous ALMACAL Band 6 map \citep[][]{Oteo2016}. The Band 8 maps of the calibrator fields with a peak flux detection at $> 4.5 \sigma$ are shown in Fig.~\ref{Chap6FigDetections}. 

We measure the flux from the primary beam corrected maps by integrating the signal in a circular aperture with a radius of $1.5$~times the synthesized beam width around the position of the peak flux determined by SExtractor. Fluxes are furthermore corrected for flux boosting. 
Additionally, we measure Band 6 and Band 7 flux densities at the position of the Band 8 detections using an aperture with a radius of $1.5$~times the beam width in Band 6 and Band 7, respectively. The multi-band flux densities of the detections are given in Table~\ref{Chap6TabDetections}.

To test whether we are missing any extended  flux in the high resolution maps, we also create maps at lower resolution of $0.8\arcsec$. We measure the flux in bright detections in the $0.8\arcsec$ maps and compare it with the flux measured in the high resolution maps. 
 We find that within the errors the two flux measurements are consistent and no correction factor needs to be applied. 
\subsection{Identifying jet emission from calibrators}

Our aim is to determine reliable 650$\mu$m number counts and for this it is necessary to identify submillimetre detections which are actually jet emission related to the calibrator. We do this by considering the submillimetre SED, the geometry of the detections and by examining radio maps of the calibrators.

For the SEDs we calculate two flux density ratios: between Bands 8 and 6 and between Bands 8 and 7. Jet emission should have a synchrotron spectrum which increases with decreasing wavelength. The emission from dust, on the other hand, is modified black body radiation which decreases with decreasing wavelength in the ALMA bands. To be conservative, we classify emission where either flux density ratio ($S_{650 \mu {\rm m}}/S_{1.2 {\rm mm}}$ or $S_{650 \mu {\rm m}}/S_{870 \mu {\rm m}}$) \la 1 as a jet. This leads to 11 jet identifications (see Table~\ref{Chap6TabDetections}). Interestingly there are reported DSFGs at $z > 4—5$ with increasing flux densities with increasing wavelength resembling jet emission \citep{Ivison2016,Riechers2017} . However, given the likely rarity of such sources (the redshift distribution of DSFGs  spans $z = $\,1--3, \citealt{Chapman2005, Dudzeviciute2020}, observed at similar wavelengths) we consider this possibility fairly unlikely.

Now considering the geometry, we identify seven fields with pairs of submillimetre detections that are diametrically opposite each other along a line which passes trough the calibrator. Such an alignment is very unlikely to happen by chance and is strongly indicative of either gravitational lensing \citep[as possible in the case of J1058+0133][]{Oteo2017} or of jet emission. We therefore consider all submillimetre detections associated with the remaining aligned pairs (J0522-3627, J1512+0203, J2056-4714, J2102+0341, J2226+0052 and J2258-2758) to likely be part of jets (\ref{Chap6TabDetections}). This adds an additional six sources to the list of jets (six are already included based on their SED) giving 17 in total.

As a check on all identifications of jet emission based on SED or alignment, we have examined radio images made ourselves from archival Karl G. Jansky Very Large Array (VLA) data, or from published maps, particularly the Australia Telescope Compact Array (ATCA) images of \citet{Marshall2005}. In most cases, the submillimetre source is indeed detected and seen to be part of a jet. Where we are not able to identify a submillimetre source with a radio jet we consider it a \textit{potential} jet. The sources are labelled in Fig.~\ref{Chap6FigDetections} as being either a jet or a potential jet.

Two fields are noteworthy in that the radio jets have a different direction to that of the aligned submillimetre pair. These are J2056-4714 and J2101+0341 and in both cases the jet directions are shown in Fig.~\ref{Chap6FigDetections}. The brighter of the two submillimetre detections in J2101+0341 is also the only one of the jet identifications which has a purely dusty spectrum i.e.\ it declines in brightness with wavelength in the ALMA bands. Given that blazar spectra with $S_{650 \mu {\rm m}}/S_{1.2 {\rm mm}} > 1$ have been reported \citep[e.g.][]{PlanckCollaboration2011, PlanckCollaboration2018} this in itself does not rule out a jet origin, but the inconsistent orientation of the radio and submillimetre structures is puzzling. However, to be conservative, we retain the identification of these four submillimetre sources as potential jets.

\begin{figure*}
\centering
\includegraphics[width = 0.49\linewidth, trim = 0 15 0 0, clip]{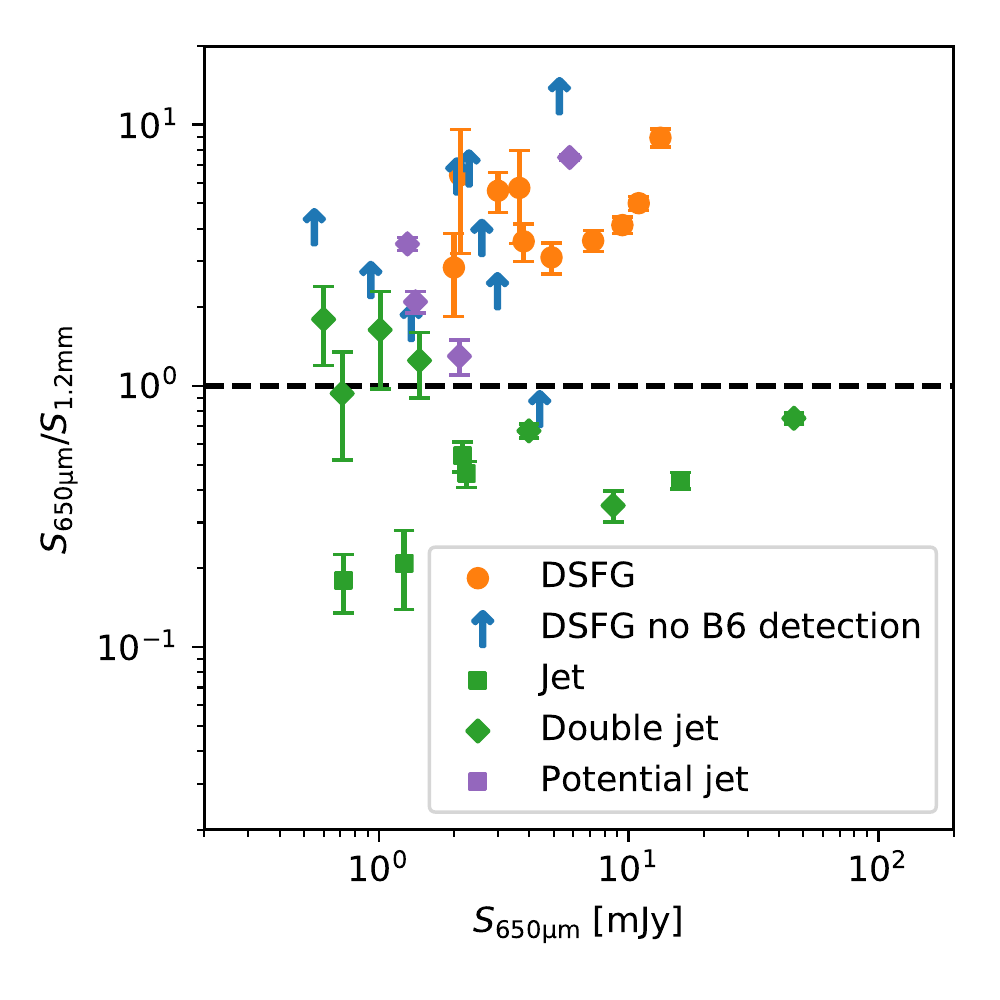}
\includegraphics[width = 0.49\linewidth, trim = 0 0 0 30, clip]{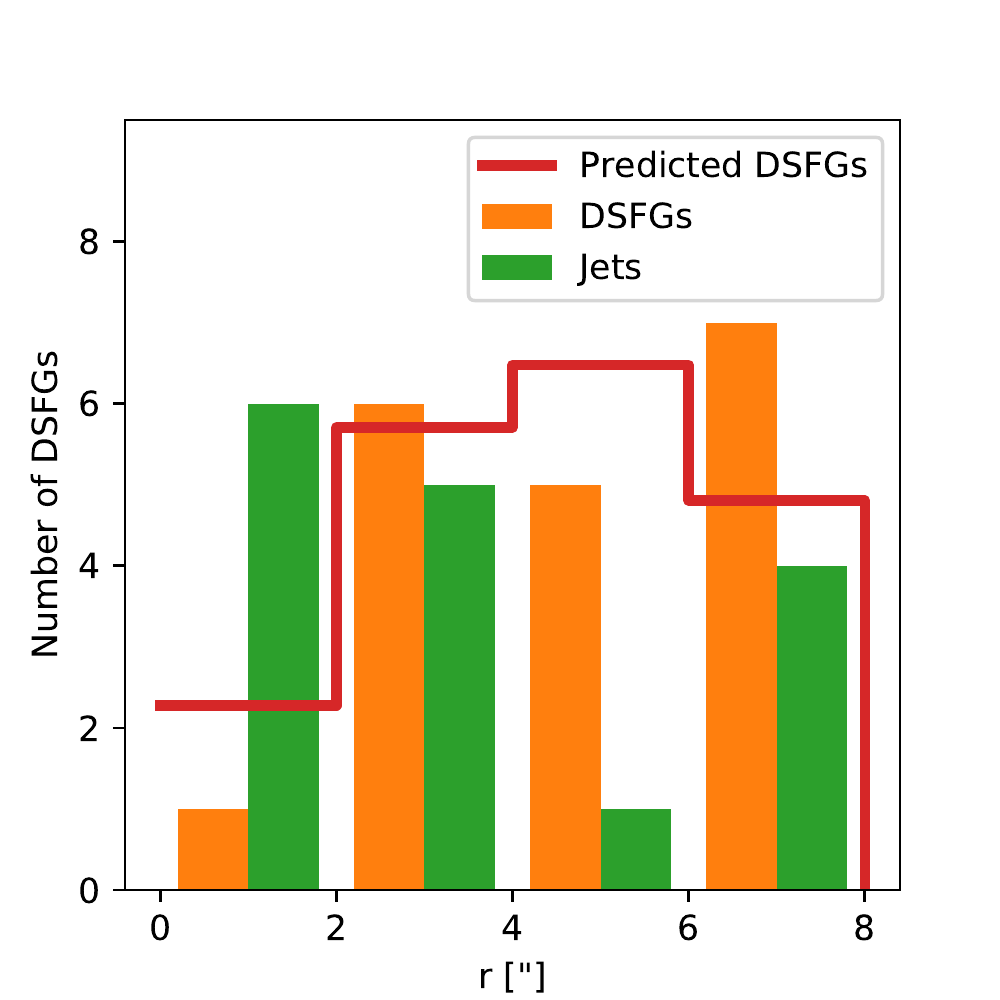}
\caption{\textit{Left panel:} Submillimetre colours of continuum objects as a function of the $650 \mu {\rm m}$ flux. All aligned double detections (except the known DSFGs in the field of J1058+0133) are classified as jets. We find that applying a colour cut of $S_{650 \mu {\rm m}}/S_{1.2 {\rm mm}} < 1$ includes all sources with a jet-like colour. Jets identified by the alignment of two continuum sources and the central calibrator ar marked with diamonds. Four of these are above the nominal threshold. We conservatively exclude these detections from further analysis. \textit{Right panel:} Number of Band 8 detections in radial annuli around the calibrator position. We compare the radial distribution of DSFGs and jets (definite and potential) with the predicted radial distribution based on the area in the radial annulus, the corresponding sensitivity and the number counts of DSFGs. We find no indication of clustering of DSFGs around the calibrator.\label{FigRadialDist}\label{FigSingDouble}}
\end{figure*}

Nine sources detected in Band 8 are not detected in Band 6 and five sources detected in Band 8 are not detected in Band 7. This could be due to the shallow Band 6 or 7 observations as indicated in Fig.~\ref{FigSingDouble}. Another explanation might be the increased cosmic microwave background (CMB) temperature at higher redshift ($z \leq 5$). As outlined by \citet{daCunha2013} the higher CMB temperatures at high redshift can result in a dimming of the long wavelength flux densities compared to short wavelength flux densities. The Band~6/7 non-detections could therefore partly trace a high redshift population of DSFGs. We cannot distinguish between the two scenarios based on the currently available ALMACAL Band 6 and 7 observations. 

In addition to the clear detections in both Band 6/7 and Band 8, we also identify a number of detections in Band 6/7 that do not have a corresponding Band 8 detection. These are expected based on the likely SED and the field of view in Band 8. These will be the subject of a future project.

\subsection{Clustering of Sources}

The ALMA calibrators are predominantly blazars \citep{Bonato2018}. These galaxies are radio bright because the line of sight coincides with the direction of the jet and not because they are particularly radioluminous or massive \citep{DeBreuck2002, Seymour2007}. However, most DSFGs are at z $\gtrsim$ 1 \citep{Dudzeviciute2020} while the calibrators are mostly below z $\sim$ 1 with a tail to z $\sim$ 3 (see Table 1). Therefore, it is less likely that the calibrator and the DSFGs are physically associated. Although we caution that it is also possible that we are biased towards higher number counts due to lensing of the DSFG by the blazar host galaxy.

Both clustering and lensing are expected to result in overdensities of sources around the calibrators and hence to search for evidence of these biases we investigate the radial distribution of candidate DSFG around the calibrators. We calculate the number of DSFG detections in radial annuli and compare these with predictions based on the area in the annuli, the sensitivity of the observations and the predicted number counts from \citet{Bethermin2017}. The expected radial distribution is calculated for individual maps. We determine the mean sensitivity per annulus in a map and derive the expected number of DSFGs in the given annulus. The sum of the radial distributions in all maps is shown in Fig.~\ref{FigRadialDist}. We find that our detected sample of DSFGs is not measurably clustered around the central calibrator and therefore we conclude that clustering is not strongly affecting our results.

\subsection{Assessing cosmic variance effects}

Calculating number counts at short wavelengths is also challenging due to the decreasing FOV with decreasing observing wavelength which can make a survey susceptible to cosmic variance effects. As the survey volume decreases small-scale inhomogeneities can start to dominate over Poissonian variations. We follow the description by \citet{Driver2010} to estimate to cosmic variance of the ALMACAL survey at $650 \mu$m. We conservatively assume a median redshift of the sources of 1 \citep{Lim2019}. The radial depth is assumed to be $z = 0 - 2$, which corresponds to $\sim 16$~Gpc. We calculate the cosmic variance for the full survey volume as well as only for the deepest maps. For the deep maps, we include only sight lines with a central r.m.s. of $< 300 \mu$Jy and take only half of the nominal search radius into account. This results in 62 sight-lines. In both cases the cosmic variance is $ \ll 5$ per cent.

\section{Results and Discussion}

\subsection{Number Counts}

\begin{figure*}
\centering
\includegraphics[width = 0.49\linewidth]{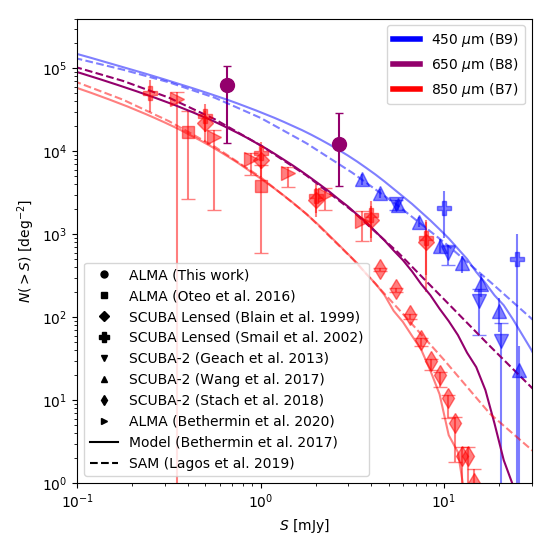}
\includegraphics[width = 0.49\linewidth]{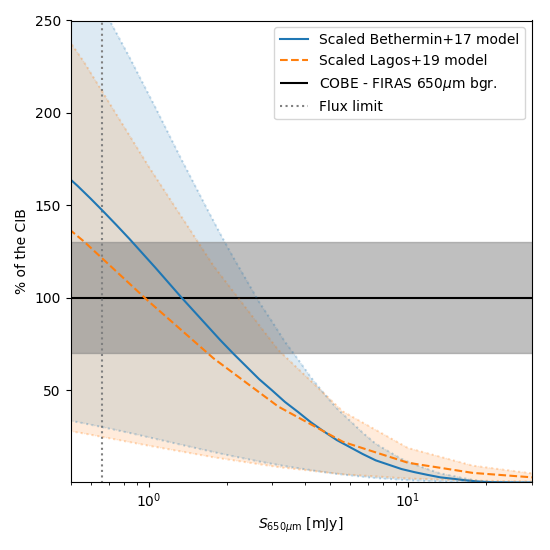}
\caption[Cumulative number counts of DSFGs at $650\mu$m.]{\textit{Left panel:}Cumulative number counts of DSFGs at $650 \mu$m (in ALMA Band 8, at 440GHz, purple). For comparison we show number counts from observations at different wavelengths as well as preditions from an empirical and semi-analytical model \citep{Bethermin2017, Lagos2019}. The number counts reported by \citet{Geach2013} are shifted by $-2.5$mJy matching the binning of the other data. \textit{Right panel:} The integrated surface brightness of the $650\mu$m emitters relative to the CIB measured by {\it COBE}-FIRAS at $650\mu$m \citep{Fixsen1998}. The models from \citet{Bethermin2017} and \citet{Lagos2019} scaled to the number counts derived in this work amount to $\simeq 100$~per cent of the cosmic infrared background at the lowest flux density observed in this study (indicated by the dotted line). The error range for the two models reflects the scaling of the models to the upper and lower limits of the number counts on the left. We conclude that we have identified the bulk of the population contributing to the EBL at $650\mu$m.\label{Chap6FigIntegratedIntensityComCIB}\label{Chap6FigNumberCounts}}
\end{figure*}

Here, we present the cumulative number counts derived from our ALMACAL Band 8 detections. In summary we have classified 21 sources as likely DSFGs, seven potential jets and ten jets. Number counts are calculated using the 21 likely DSFGs.
These are the shortest wavelength number counts yet derived from interferometric observations.

A galaxy $i$ contributes to the cumulative number counts as follows:

\begin{equation}
N_i(S_i) = \frac{1 - f_{{\rm SP} (S_i)}}{C(S_i) \times A(S_i)},
\end{equation}

\noindent where $S_i$ is the flux density of the source $i$, $f_{{\rm SP} (S_i)}$ is the fraction of spurious sources at $S_i$, $C(S_i)$ is the survey completeness at $S_i$ and $A(S_i)$ is the effective area covered by the survey at $S_i$. Using our multi-band data we have excluded all sources with Band 8 to Band 6 of Band 7 flux density ratios indicative of jets. The effective area and completeness are taken from Sections~\ref{Chap6SecCompleteness} and~\ref{Chap6SecArea}, respectively. To calculate the cumulative number counts we sum over all galaxies with flux densities higher than a given value:

\begin{equation}
N(> S) = \sum_{S_i > S} \frac{1 - f_{{\rm SP} (S_i)}}{C(S_i) \times A(S_i)}.
\end{equation}

To calculate the errors on the cumulative number counts we combine bootstrapping with Poissonian errors. First we assign random fluxes to all detections within the uncertainties quoted in Table~\ref{Chap6TabDetections}. This is done 1000 times to derive alternative realizations of the number counts. The bootstrapping error is the standard deviation in the 1000 realizations of the number counts. Second we determine the Poissonian errors for a 1$\sigma$ confidence level given the number of high and low flux sources using the tables provided by \citet{Gehrels1986}. Furthermore, we account for misidentified sources from jets using the Poissonian error of the number of potential jets. This error encompasses residual contamination by single-sided jets with flux ratios $> 1$.
The total error is the quadratic sum of the bootstrap and Poissonian errors. 
Results are shown in Fig.~\ref{Chap6FigNumberCounts} and listed in Table~\ref{Chap6TabNUmberCounts}. 

The derived number counts follow the expected trend of increasing number counts with decreasing observing wavelength. The number counts at $650 \mu$m presented here fall in between those at $450 \mu$m and $850 \mu$m presented in previous works. At the same time we are reaching lower flux densities than previous works at $450 \mu$m by almost one order of magnitude. At this shorter wavelength we are probing emission closer to the peak of the dust emission in the infrared which is directly related to the obscured star formation. 
We show in Fig.~\ref{Chap6FigIntegratedIntensityComCIB} that the Band 8 number counts are also consistent with the predicted number counts based on semi-empirical \citep{Bethermin2017} and semi-analytic galaxy formation models \citep{Lagos2019}. In the Simulated Infrared Dusty Extragalactic Sky (SIDES, \citealt{Bethermin2017}) a dark-matter simulation is populated using empirically-calibrated relations such as the stellar mass versus halo mass relation and the main sequence of star-forming galaxies (see method in \citealt{Bethermin2013}). It relies on recent SIDES templates measured using {\it Herschel} stacking \citet{Bethermin2015}. This simulation accurately reproduces the properties of infrared and submillimetre sources taking into account the non-negligible effects caused by the resolution of instruments. \citet{Lagos2019} presented predictions for the FUV-to-FIR emission of galaxies for the physical semi-analytic model of galaxy formation Shark \citep{Lagos2018}. Unlike previous work, they adopted a universal initial mass function to show that reproducing the panchromatic emission of galaxies was possible.

We assess whether the derived number counts are sensitive to the classification of the potential jets. We calculate the number counts with and without these seven sources and find that the number count in the upper flux bin changes by less than 5 per cent while the number count in the lower flux bin changes by up to 40 per cent.



\begin{table}
\centering
\caption{Cumulative $650\mu$m number counts.\label{Chap6TabNUmberCounts}}
\begin{tabular}{l l}
\hline
$S$ [mJy] & $\log N(> S)$ [deg$^{-2}$]\\
\hline
0.67 & $4.8\pm 4.7$\\
2.50 & $4.1^{+3.9}_{-4.2}$\\
\hline
\end{tabular}
\end{table}

\subsection{Resolving the 650$\mu$m Background Light}

We assess what fraction of the cosmic infrared background (CIB) is resolved by our observations. Due to the limited number of detections we do not constrain the shape of the number counts. Therefore, we scale the semi-empirical and semi-analytical model predictions from \citet{Bethermin2017} and \citet{Lagos2019} to our data points. 

By integrating the scaled models we calculate the integrated surface brightness of the $650\mu$m emitters and compare this with the cosmic infrared background at the same wavelength as measured by the \textit{COBE}-Far Infrared Absolute Spectrophotometer (FIRAS) \citep[][]{Fixsen1998}. 
The authors find a CIB flux density of $I_{\nu} (650 \mu {\rm m}) = (0.22 \pm 0.07) {\rm MJy \; sr^{-1}}$. At the lowest observed flux density we find an integrated flux density of point sources detected at $650 \mu$m of $I_{\nu} (650 \mu {\rm m}) = 0.34 ^{+0.26}_{-0.27} {\rm MJy \; sr^{-1}}$ and $I_{\nu} (650 \mu {\rm m}) = 0.26 ^{+0.23}_{-0.22} {\rm MJy \; sr^{-1}}$ for the two models, respectively (see Fig.~\ref{Chap6FigIntegratedIntensityComCIB}). Therefore, we are resolving $150 \pm 120$ or $130 \pm 100$ per cent for the two models, respectively. We conclude that the $650 \mu$m ALMACAL observations are deep enough to resolve the majority of the cosmic infrared background at $650 \mu$m. 

\section{Summary and Conclusion}

In this paper we present the first short wavelength number counts at $650\mu$m free of blending and cosmic variance. We use observations from the ALMACAL survey until December 2018. In 81 fields 21 DSFGs were detected at a detection threshold of $4.5 \sigma$ reaching flux densities as low as 0.66mJy, roughly an order of magnitude lower than previous \textit{Herschel} and SCUBA-2 surveys. We combine the detections in Band 8 with observations of the same fields in Band 6 and 7 from ALMACAL to identify and remove jets associated with the calibrators. Of the 21 DSFGs detected at $650\mu$m 10 are also detected at 1.2mm and 11 are detected at 870$\mu$m. We carefully identify jet emission using submillimetre colours as well as radio emission maps. 
We do not find a spatial correlation between the DSFG position and the position of the calibrators in the respective fields. 
The cumulative number counts follow the expected trend of increasing number counts with decreasing observing wavelength. They are also consistent with predictions from semi-empirical and semi-analytical models \citep{Bethermin2017, Lagos2019}. These number counts at shorter infrared wavelengths probe the dust emission closer to the peak of the dust SED. With this work we approach a regime that so far has only been accessible with low resolution observations or with the aid of lensing. Furthermore, we reach flux densities sensitive enough to resolve $150 \pm 120$ or $130 \pm 100$ per cent of the cosmic infrared background at $650 \mu$m for the two models, respectively. This is a significant improvement over the $24 - 33$~per cent previously reached at $450 \mu$m with SCUBA-2 \citep{Wang2017}. Our finding is comparable with that of \citet{Chen2013}, who resolved 90 per cent of the EBL at $450 \mu$m tracing flux densities down to 1~mJy. A larger survey area would be beneficial to pin down the exact shape of the number counts at this wavelength. However, this study includes the largest available ALMA dataset corresponding to 112 hours of observing time obtained during six years of ALMA operation. Even using a dedicated Large Program could only double the the total survey area at similar depth. Objects with flux >0.7 mJy make up most of the EBL at $650 \mu$m. We expect more numerous fainter objects likely contribute only a small fraction to the EBL.

\section*{Acknowledgements}

AK thanks Loretta Dunne and Simon Morris for interesting discussions and comments that helped to improve the paper. 
AK gratefully acknowledges support from the STFC grant ST/P000541/1 and Durham University and the Independent Research Fund Denmark via grant number DFF 8021-00130
CP thanks the Alexander von Humboldt Foundation for the granting of a Bessel Research Award held at MPA. 
IRS acknowledges support from STFC (ST/P000541/1). 
C.C.C. acknowledges support from the Ministry of Science and Technology of Taiwan (MOST 109-2112-M-001-016-MY3). 
RD thanks the Alexander von Humboldt Foundation for support. 
ALMA is a partnership of ESO (representing its member states), NSF (USA), and NINS (Japan), together with NRC (Canada), NSC and ASIAA (Taiwan), and KASI (Republic of Korea), in cooperation with the Republic of Chile. The Joint ALMA Observatory is operated by ESO, AUI/NRAO, and NAOJ. 
This paper makes use of the following ALMA data: \\ 
ADS/JAO.ALMA
\#2013.1.00020.S
\#2013.1.00048.S
\#2013.1.00061.S
\#2013.1.00071.S
\#2013.1.00139.S
\#2013.1.00170.S
\#2013.1.00180.S
\#2013.1.00227.S
\#2013.1.00244.S
\#2013.1.00368.S
\#2013.1.00433.S
\#2013.1.00448.S
\#2013.1.00527.S
\#2013.1.00535.S
\#2013.1.00576.S
\#2013.1.00749.S
\#2013.1.00907.S
\#2013.1.01010.S
\#2013.1.01052.S
\#2013.1.01099.S
\#2013.1.01202.S
\#2013.1.01230.S
\#2015.1.00019.S
\#2015.1.00035.S
\#2015.1.00384.S
\#2015.1.00420.S
\#2015.1.00612.S
\#2015.1.00631.S
\#2015.1.00650.S
\#2015.1.00656.S
\#2015.1.00662.S
\#2015.1.00717.S
\#2015.1.00736.S
\#2015.1.00770.S
\#2015.1.00847.S
\#2015.1.00942.S
\#2015.1.01040.S
\#2015.1.01133.S
\#2015.1.01137.S
\#2015.1.01163.S
\#2015.1.01191.S
\#2015.1.01486.S
\#2015.1.01487.S
\#2015.1.01518.S
\#2015.1.01597.S
\#2015.A.00018.S
\#2015.A.00026.S
\#2016.1.00089.S
\#2016.1.00226.S
\#2016.1.00234.S
\#2016.1.00262.S
\#2016.1.00263.S
\#2016.1.00284.S
\#2016.1.00333.S
\#2016.1.00346.S
\#2016.1.00441.S
\#2016.1.00668.S
\#2016.1.00805.S
\#2016.1.00954.S
\#2016.1.01093.S
\#2016.1.01137.S
\#2016.1.01253.S
\#2016.1.01285.S
\#2016.1.01293.S
\#2016.1.01313.S
\#2016.1.01399.S
\#2016.1.01613.S
\#2017.1.00005.S
\#2017.1.00025.S
\#2017.1.00043.S
\#2017.1.00078.S
\#2017.1.00118.S
\#2017.1.00190.S
\#2017.1.00201.S
\#2017.1.00239.S
\#2017.1.00337.S
\#2017.1.00345.S
\#2017.1.00370.S
\#2017.1.00467.S
\#2017.1.00508.S
\#2017.1.00586.S
\#2017.1.00753.S
\#2017.1.00775.S
\#2017.1.00814.S
\#2017.1.00984.S
\#2017.1.01052.S
\#2017.1.01195.S
\#2017.1.01209.S
\#2017.1.01457.S
\#2017.A.00024.S
\#2017.A.00047.T
\#2018.1.00065.S
\#2018.1.00649.S

The National Radio Astronomy Observatory is a facility of the National Science Foundation operated under cooperative agreement by Associated Universities, Inc. 
This research made use of Astropy,\footnote{http://www.astropy.org} a community-developed core Python package for Astronomy \citep{astropy:2013, astropy:2018}. 




\bibliographystyle{mnras}



%

%

\bsp	
\label{lastpage}
\end{document}